\documentclass[11pt,twoside]{article}

\usepackage{asp2006}
\usepackage{epsf}
\usepackage{lscape}

\begin{document}
\title{Near-Infrared Polarimetry toward the Galactic Center}   

\author{Shogo Nishiyama\altaffilmark{1}, Motohide Tamura\altaffilmark{2},
Hirofumi Hatano\altaffilmark{3}, Saori Kanai\altaffilmark{3}, 
Mikio Kurita\altaffilmark{3}, Shuji Sato\altaffilmark{3},
Tetsuya Nagata\altaffilmark{1}}   
\altaffiltext{1}{Department of Astronomy, Kyoto University, Kyoto 606-8502, Japan} 
\altaffiltext{2}{National Astronomical Observatory of Japan, Mitaka, Tokyo 181-8588, Japan}
\altaffiltext{3}{Department of Astrophysics, Nagoya University, Nagoya 464-8602, Japan}

\begin{abstract} 
Near-infrared polarimetry of point sources
reveals the presence of a toroidal magnetic field
in the central 20' x 20' region of our Galaxy.
Comparing the Stokes parameters 
between high extinction stars and relatively low extinction ones, 
we have obtained a polarization originating from magnetically aligned dust grains 
at the central region of our Galaxy of at most 1-2 kpc. 
The derived direction of the magnetic field is 
in good agreement with that obtained from far-infrared/submillimeter observations, 
which detect polarized thermal emission from dust in the molecular clouds at the Galactic center. 
Our results show that by subtracting foreground components, 
near-infrared polarimetry allows investigation of the magnetic field structure at the Galactic center. 
The distribution of the position angles shows a peak at around 20$^{\circ}$, 
nearly parallel to the direction of the Galactic plane, suggesting a toroidal magnetic configuration.
\end{abstract}



\section{Introduction}

The magnetic field configuration at the Galactic center (GC) has been investigated
with a wide variety of methods.
Recent far-infrared (FIR) and sub-millimeter (sub-mm) polarimetric observations
point out that the magnetic field is generally parallel to the Galactic plane
\citep{Novak00,Novak03,Chuss03}.
This is contrast to the poloidal field traced by non-thermal radio filaments,
most of which align nearly perpendicular to the Galactic plane.

Previous near-infrared (NIR) polarization measurements of the GC
were discussed in terms of selective absorption by the intervening 
interstellar dust grains in the Galactic disk.
To our knowledge, no one has studied the magnetic field configuration {\it at} the GC
with NIR polarimetry.
In this paper, we present results of NIR polarimetric observations toward the GC.
We demonstrate that NIR polarization of point sources can provide information
on the magnetic field structure at the central region of our Galaxy.

\section{Observations and Data Analysis}

We observed a 20' $\times$ 20' area
centered at the position of Sgr A*
with the NIR polarimetric camera SIRPOL on the IRSF telescope.
SIRPOL consists of a single-beam polarimeter
and NIR imaging camera SIRIUS.
SIRIUS provides images of a 7\farcm7 $\times$ 7\farcm7 area of sky 
in three NIR wavebands, $J$ ($1.25\mu$m), $H$ ($1.63\mu$m),
and $K_S$ ($2.14\mu$m) simultaneously
with a scale of 0\farcs45 pixel$^{-1}$.

The Stokes parameters $I$, $Q$, and $U$ for point sources
were determined from aperture polarimetry of combined images.
Based on intensities for each wave plate angle,
we calculated the Stokes parameters $I,Q,$ and $U$ as
$I = (I_{0\fdg0} + I_{22\fdg5} + I_{45\fdg0} + I_{67\fdg5})/2$,
$Q = I_{0\fdg0} - I_{45\fdg0},$
and $U = I_{22\fdg5} - I_{67\fdg5}$.
The degree of polarization $P$ and the position angle $\theta$ were derived by 
\[P = \sqrt{(Q^2+U^2)}/I,
~~~\theta = \frac{1}{2} \arctan(U/Q). \]

We show a $K_S$-band vector map binned by 
$0\farcm5 \times 0\farcm5$ in Fig. 1.
The vectors are superposed on 
the three color ($J$, $H$, $K_S$) composite image of the same region.
At a first glance, 
most of the vectors are in order,
and are nearly parallel to the Galactic plane.
Moving north-eastward across the image,
the position angles slightly rotate clockwise.
At a few positions where the number density of stars is small,
and hence strong foreground extinction exists,
the vectors have irregular directions
particularly at the northwestern corner.
These irregularity might be explained by
the inherent magnetic field configuration in foreground dark clouds.

\begin{figure}[!ht]
  \plotone{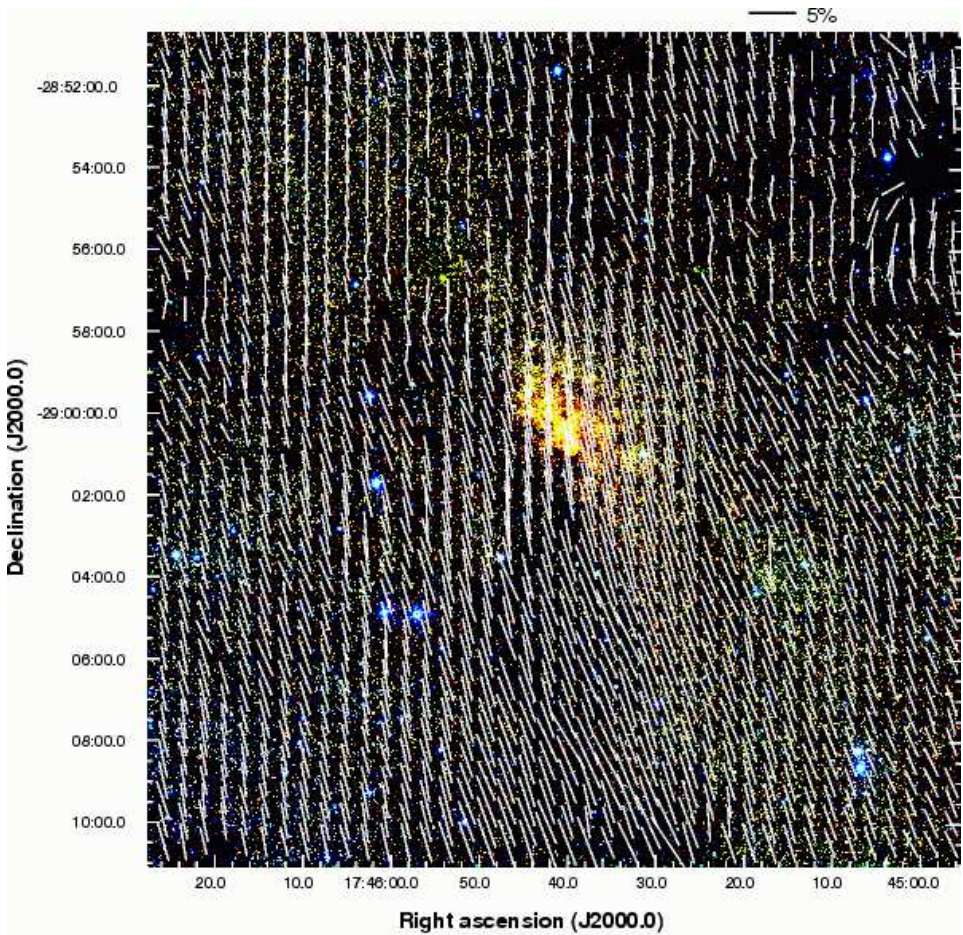}
  \caption{
    $K_S$-band polarization vector map superposed on 
    the three color ($J$, $H$, $K_S$) composite image of the Galactic center.
    The Galactic center is the bright yellow blob in the center.
    The mean $P_{K_S}$ and ${\theta}_{K_S}$ are calculated 
    for each $0\farcm5 \times 0\farcm5$ grid.
  }
\end{figure}

\section{Magnetic Field Configuration at the Galactic Center}

Polarimetric measurements of stars of different distances reveal
the three dimensional distribution of magnetic field orientations.
From the stars at the close side in the Galactic bulge 
(referred to hereafter as ``blue stars'' due to their relatively small reddening),
we can obtain $P$ and $\theta$,
which are affected mainly by interstellar dust grains in the Galactic disk.
The light from stars at the far side in the bulge
(hereafter ``red stars'')
is transmitted through the dust in the disk {\it and} the bulge.
Therefore, using both blue and red stars,
we can obtain the polarization originating in the bulge. 
The procedure is as follows.

At first, we divided the field into $10 \times 10$ sub-fields,
and made $H-K_S$ histograms for each sub-field.
Using the histograms, we evaluated a peak value of the histogram $(H-K_S)_{\mathrm{peak}}$.
For $H-K_S$ color, we divided the stars into two groups:
``blue'' and ``red'' stars in the bulge.
The ``blue'' stars are redder than $H-K_S=1.0$ and 
bluer than ${(H-K_S)}_{\mathrm{peak}}$.
The stars with $H-K_S > {(H-K_S)}_{\mathrm{peak}}$
are selected as ``red'' stars.

Next, $Q/I$ and $U/I$ histograms in the $K_S$ band were constructed
for the blue and red stars in each sub-field.
We calculated their means as
$<Q/I>_{\mathrm B}$, $<Q/I>_{\mathrm R}$,
$<U/I>_{\mathrm B}$, and $<U/I>_{\mathrm R}$.
We then obtained $P$ and $\theta$
for ``red minus blue'' components
using the following equations \citep{Goodrich86}:
\[
P_{\mathrm {R-B}} = \sqrt{ \left( \left< \frac{Q}{I} \right>_{\mathrm R} - \left< \frac{Q}{I} \right>_{\mathrm B} \right)^{2} 
  + \left( \left< \frac{U}{I} \right>_{\mathrm R} - \left< \frac{U}{I} \right>_{\mathrm B}  \right)^{2} }, 
\]
\[
\theta_{\mathrm {R-B}} = \frac{1}{2} \arctan \left[ 
\Big( \left< \frac{U}{I} \right>_{\mathrm R} - \left< \frac{U}{I} \right>_{\mathrm B} \Big) \bigg/
\Big( \left< \frac{Q}{I} \right>_{\mathrm R} - \left< \frac{Q}{I} \right>_{\mathrm B} \Big) \right].
\]
The errors of $<Q/I>$ and $<U/I>$ were calculated from the standard error on the mean
$\sigma/\sqrt{N}$ of the $Q/I$ and $U/I$ histograms, where $\sigma$ is the standard deviation 
and $N$ is the number of stars (see Nishiyama et al. 2009, for more detail).

We show a vector map for $P_{\mathrm {R-B}}$ and $\theta_{\mathrm {R-B}}$ in Fig. 2.
The average of $P_{\mathrm {R-B}}$ and $\theta_{\mathrm {R-B}}$ are
obtained as 0.85 \% and 16\fdg0, only for grids where the polarization
is detected with $P_{\mathrm {R-B}}/\delta P_{\mathrm {R-B}} \geq 2$.
The histogram of $\theta_{\mathrm {R-B}}$ has a peak at $\sim 20^\circ$,
which roughly coincides with the angle of the Galactic plane.
A similar result is also obtained for the $H$-band polarization.
This coincidence shows the basically toroidal geometry of the magnetic field.

\section{Discussion}

The direction of the magnetic field at the GC has been investigated 
from polarized dust emission in the FIR and sub-mm wavelengths.
As seen in Fig. 2,
the magnetic field configuration we obtained at the GC 
shows a good agreement globally with those obtained by
\citet{Dotson00}, \citet{Novak00}, and \citet{Chuss03},
which are the highest angular resolution polarimetry data sets in the FIR/sub-mm wavelengths.
The polarized FIR/sub-mm emission comes from molecular clouds,
which are known to be located in the GC.
Therefore we conclude that the position angles derived from our NIR polarimetry
represent the direction of the magnetic field {\it in} the GC.

\begin{figure}[!ht]
  \plotone{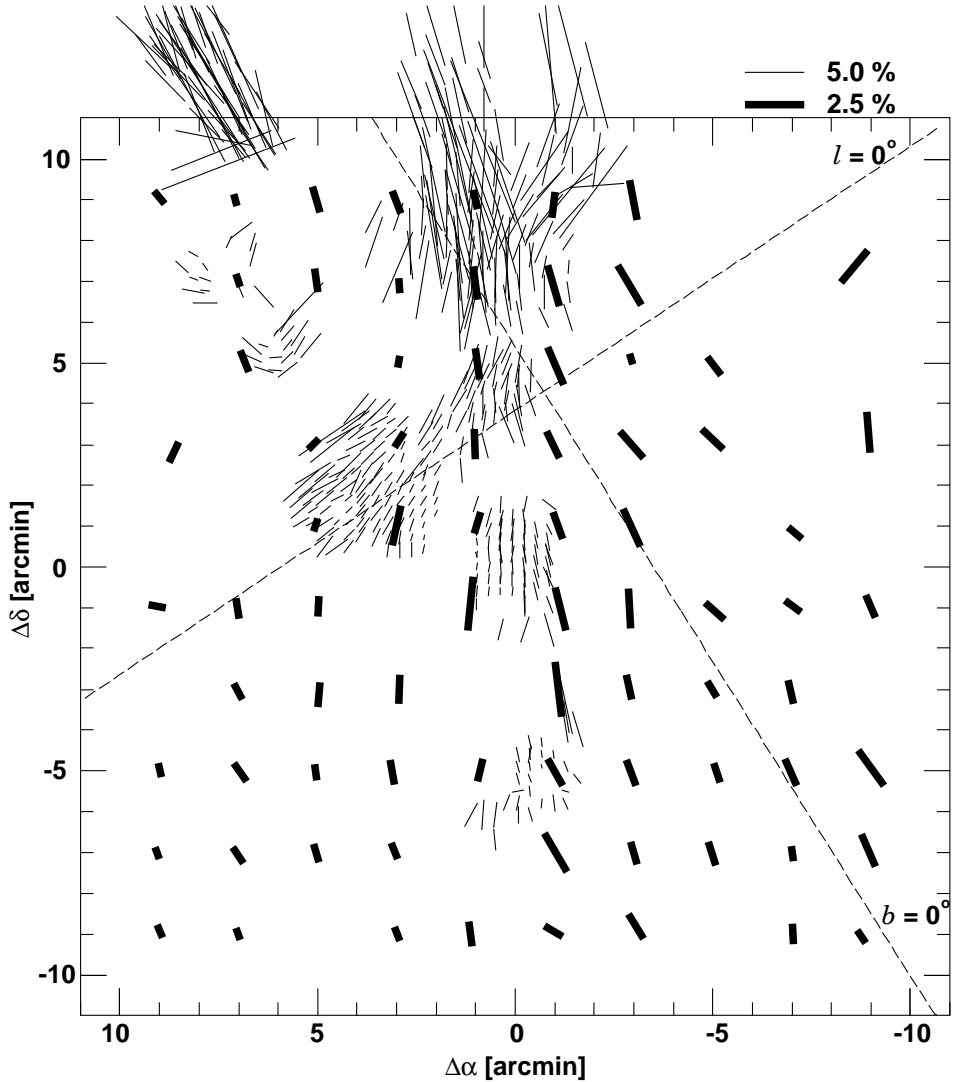}
  \caption{$K_S$-band vector map derived from the Galactic center component 
    ($P_{\mathrm {R-B}}$ \& $\theta_{\mathrm {R-B}}$, thick bars).
    The length of the bars is proportional to the measured degree of polarization,
    and their orientation is drawn parallel to the inferred magnetic field direction.
    The polarization map derived from FIR/sub-mm observations (thin bars) is also shown.
    The data sets of FIR/sub-mm wavelengths are from 
    60 $\mu$m \& 100 $\mu$m polarimetry by \citet{Dotson00},
    and 350 $\mu$m polarimetry by \citet{Novak00} and \citet{Chuss03}.
  }
\end{figure}

From $H-K_S$ color differences between peaks in the $H-K_S$ histograms
and mean colors of the blue and red stars,
we tried to estimate the depth of the region
where we have mapped the magnetic field configuration.
Using $A_{K_S}/E_{H-K_S}=1.44$ \citep{Nishi06}
and the model of dust distribution by \citet{Davies97},
we obtained the average distances of 0.5 kpc from the GC for the blue stars,
and 1.0 kpc for the red stars.
This suggests that the polarization shown in Fig. 2 occurs
between the average distances of $(R_0 - 0.5)$ kpc and $(R_0 + 1.0)$ kpc from the Sun,
arising probably from the central 1$-$2 kpc region of our Galaxy
(where $R_0$ is the distance between the GC and the Sun).

We have shown that the polarization of starlight can be 
a probe of the magnetic field near the GC.
\citet{Morris98} enumerated five different ways
in which the magnetic field near the GC has been studied:
morphology, polarization angle, Faraday rotation of the radio continuum,
Zeeman effect, and polarized dust emission in FIR/sub-mm wavelengths.
The wide field-of-view of the NIR polarimeter SIRPOL,
and the statistical treatment of tens of thousands of stars
enable us to study the magnetic field near the GC;
that is, the NIR polarimetry of starlight is a new way 
to investigate the magnetic field {\it in} the GC.

NIR polarimetry has the advantage of
providing information about the magnetic field
at locations where FIR/sub-mm emission is weak.
NIR polarization of starlight is attributed 
to extinction along the line of sight by aligned dust grains,
while FIR/sub-mm polarization is due to emission from the aligned dust.
Hence, NIR polarimetry can investigate the magnetic field in regions
where FIR/sub-mm polarimetry is absent, if background stars exist.
This advantage is clearly shown in Fig. 2.
We have detected polarization at positions where thin bars are not shown.
The distribution of the position angles in most of the observed regions
including such low emission regions
shows a globally toroidal magnetic configuration at the GC.


\acknowledgements 
SN is financially supported by the Japan Society for the Promotion of Science (JSPS) 
through the JSPS Research Fellowship for Young Scientists.
This work was also supported by KAKENHI,
Grant-in-Aid for Young Scientists (B) 19740111,
and Grant-in-Aid for Scientific Research (A) 19204018.


\end{document}